\documentstyle[aps,preprint,floats,tighten]{revtex}
\input{psfig.tex}
\begin{document}
\draft
\preprint{\vbox{\hbox{CU-TP-703}
                \hbox{SU-4240-614}
                \hbox{POP-629}
                \hbox{CfA-4136}
                \hbox{astro-ph/9507080}
}}

\title{\vskip 1in
Weighing the Universe with the Cosmic Microwave
Background}

\author{Gerard Jungman\footnote{jungman@npac.syr.edu}}
\address{Department of Physics, Syracuse University,
Syracuse, New York~~13244}
\author{Marc Kamionkowski\footnote{kamion@phys.columbia.edu}}
\address{Department of Physics, Columbia University,
New York, New York~~10027}
\author{Arthur Kosowsky\footnote{akosowsky@cfa.harvard.edu}}
\address{Harvard-Smithsonian Center for Astrophysics,
60 Garden Street, Cambridge, Massachusetts~~02138
\\and\\
Department of Physics, Lyman Laboratory, Harvard University,
Cambridge, Massachusetts~~02138}
\author{David N.~Spergel\footnote{dns@astro.princeton.edu}}
\address{Department of Astrophysical Sciences, Princeton University,
Princeton, New Jersey~~08544}
\date{July, 1995}
\maketitle

\vfil\eject

\begin{abstract}
Variations in $\Omega$, the total density of the Universe, leave
a clear and distinctive imprint on the power spectrum of temperature
fluctuations in the cosmic microwave background (CMB). This signature
is virtually independent of other cosmological parameters or
details of particular cosmological models. We evaluate the precision
with which $\Omega$ can be determined by a CMB map as a function of
sky coverage, pixel noise, and beam size.
For example, assuming only that the primordial density
perturbations were adiabatic and with no prior information
on the values of any other cosmological parameters,
a full-sky CMB map at $0.5^\circ$
angular resolution and a noise level of $15\,\mu{\rm K}$ per
pixel can determine $\Omega$ with a variance of 5\%.
If all other cosmological parameters are fixed, $\Omega$ can be
measured to better than 1\%.
\end{abstract}

\pacs{98.70.V, 98.80.C}

Determination of the geometry of the Universe remains perhaps
the most compelling problem in cosmology. Alternatively
stated, what is the mean total energy density of the Universe?
The answer to this question will reveal the ultimate fate of
the Universe. If the density $\Omega$ (in units of the
critical density $\rho_c = 3 H_0^2/8\pi G$, where
$H_0$ is the Hubble constant) is greater than unity, the
Universe is closed and will eventually recollapse; if it is
less than unity, the Universe will expand forever; and if
$\Omega=1$, the expansion will asymptotically decelerate
to zero.

Theoretical considerations favor a critical ($\Omega=1$) Universe,
and inflation provides a generic mechanism for obtaining
$\Omega=1$. However, luminous matter provides less than
one percent of this mass. Various inferences of $\Omega$
by dynamical means have hinted at substantial amounts of
unseen mass, but most traditional methods of determining
$\Omega$ are plagued by systematic uncertainties. Furthermore,
virtually all dynamical methods of obtaining $\Omega$ give
the mean density in only nonrelativistic matter, and thus
cannot discriminate between an open Universe and a flat Universe that
is dominated by  vacuum energy (i.e., a cosmological constant).

Recently, it was proposed that
temperature anisotropies in the cosmic microwave background
(CMB) might be used to determine the geometry of the Universe
\cite{kamspergelsug}. Features (known as ``Doppler
peaks,'' or more accurately as acoustic peaks) in the CMB
angular power spectrum result from acoustic oscillations
in the photon-baryon fluid before the photons decouple.
The characteristic wavelength of these fluctuations is the
sound horizon at decoupling (the distance an acoustic disturbance
propagates from $t=0$ until decoupling), which subtends an angular
scale on the sky today of $\theta\simeq 1^\circ \Omega^{1/2}$.
The dependence on $\Omega$ arises directly from the geometry
of the Universe, and this angular scale is
largely independent of other cosmological parameters.
Thus, the location of the first Doppler peak
provides a robust determination of $\Omega$.
A CMB map with fine angular resolution also constrains the other
cosmological parameters by measuring the angular locations
and amplitudes of the higher Doppler peaks.

In this paper, we evaluate the precision with which $\Omega$ can
be determined with high-resolution CMB maps \cite{cobe2}. We work within
the context of models with adiabatic primordial density
perturbations, although similar arguments apply to isocurvature
models as well \cite{ct}, and we expect the power
spectrum to distinguish clearly the two classes of models.
We also briefly consider what information on other cosmological
parameters the CMB can provide.

A given cosmological theory makes a statistical prediction
about the distribution of CMB temperature fluctuations,
expressed by the angular power spectrum
\begin{equation}
C(\theta) \equiv \left\langle {\Delta T({\bf\hat m})\over T_0}
                         {\Delta T({\bf\hat n})\over T_0}
			 \right\rangle_{ {\bf\hat m}\cdot{\bf\hat n} =
			 \cos\theta}
          \equiv \sum_\ell {2\ell+1\over 4\pi} C_\ell
                     P_\ell(\cos\theta),
\label{powerspectrum}
\end{equation}
where $\Delta T({\bf\hat n})/T_0$ is the fractional temperature
fluctuation in the direction $\bf\hat n$, $P_\ell$ are the Legendre
polynomials, and the brackets represent an ensemble average over
all observers and directions. The mean CMB temperature is
$T_0=2.726 \pm 0.010 \,{\rm K}$ \cite{firas}.
Since we can only observe a single microwave
sky, the observed multipole moments $C_\ell^{\rm obs}$ will be
distributed about the mean value $C_\ell$ with a ``cosmic variance''
$\sigma_\ell \simeq \sqrt{2/(2\ell+1)}C_\ell$; no measurement can
determine the $C_\ell$ to better accuracy than this variance.

We consider an experiment which maps a fraction $f_{\rm sky}$
of the sky with a gaussian beam with full width at half maximum
$\theta_{\rm fwhm}$ and a pixel noise
$\sigma_{\rm pix} = s/\sqrt{t_{\rm pix}}$, where $s$ is the detector
sensitivity and $t_{\rm pix}$ is the time spent observing each
$\theta_{\rm fwhm}\times\theta_{\rm fwhm}$ pixel. We adopt the
inverse weight per solid angle,
$w^{-1}\equiv (\sigma_{\rm pix}\theta_{\rm fwhm}/T_0)^2$,
as a measure of noise that is pixel-size independent\cite{knox}.
Current state-of-the-art detectors achieve sensitivities of
$s=200\,\mu {\rm K}\,\sqrt{\rm sec}$, corresponding to an inverse
weight of $w^{-1}\simeq 2\times 10^{-15}$ for a one-year experiment.
Realistically, however, foregrounds and other systematic effects may
increase the noise level; conservatively, $w^{-1}$ will likely fall
in the range $(0.9-4)\,\times\,10^{-14}$.
Treating the pixel noise as gaussian and ignoring any
correlations between pixels, estimates of
$C_\ell$ can be approximated as normal distributions with
a variance (modified from Ref.~\cite{knox})
\begin{equation}
\sigma_\ell = \left[{2\over (2\ell +1)f_{\rm sky}}\right]^{1/2}
              \left[C_l + (w f_{\rm sky})^{-1} e^{\ell^2\sigma_b^2}\right].
\label{variance}
\end{equation}

Given a spectrum of primordial density perturbations, the $C_\ell$
are obtained by solving the coupled equations for the evolution of
perturbations to the spacetime metric and perturbations to the
phase-space densities of all particle species in the Universe.
We consider models with initial adiabatic density perturbations
filled with
photons, neutrinos, baryons, and collisionless dark matter; this
includes all inflation-based models.
We begin with approximate analytic solutions for the scalar
\cite{hu}\ and tensor \cite{wang}\ metric perturbations.  Our
calculation includes polarization \cite{zh}, scale dependence
of the initial perturbation spectrum \cite{kt}, and the large-angle
integrated Sachs-Wolfe effect from a cosmological constant \cite{kofman}.
To a good approximation, reionization can be parameterized by
the optical depth $\tau$ to the surface of last scatter
\cite{kamspergelsug}; anisotropies on scales much smaller
than the horizon at reionization are suppressed by $e^{-2\tau}$
while those on larger scales are unaffected.
The geometry of the Universe is then accounted for by shifting
the moments, $C_\ell(\Omega) = C_{\ell\Omega^{1/2}}(\Omega=1)$ \
\cite{kamspergelsug},
and approximating the large-angle integrated Sachs-Wolfe effect
in an open universe \cite{kamspergel}. We do not here account for
massive neutrinos (hot-- or mixed-- dark-matter models), but the
power spectrum is altered only slightly by trading some of the
nonrelativistic matter for neutrinos and our results should
be unchanged\cite{mdm}.

The CMB power spectrum depends upon many parameters. In the present
analysis we include the following set: the total density $\Omega$;
the Hubble constant, $H_0 = 100\;h\,{\rm km\,sec^{-1}\,Mpc^{-1}}$;
the density of baryons in units of the critical density, $\Omega_b h^2$;
the cosmological constant in units of the critical density, $\Lambda$;
the power-law indices of the initial scalar- and tensor-perturbation
spectra, $n_S$ and $n_T$; the amplitudes of the scalar and tensor
spectra, parameterized by $Q$, the total CMB quadrupole moment, and
$r=Q_T/Q_S$, the ratio of the tensor and scalar quadrupole moments;
the optical depth to the surface of last scatter, $\tau$; the
deviation from scale invariance of the scalar perturbations,
$a_{\rm run}\equiv dn/d\ln k$; and the effective number of
light-neutrino species at decoupling, $N_\nu$.
Thus for any given set of cosmological parameters
${\bf s}=\{\Omega,\Omega_b h^2,h,n_S,\Lambda,r,n_T,a_{\rm
run},\tau,Q,N_\nu\}$, we can calculate the mean multipole moments
$C_\ell({\bf s})$.

We now wish to determine the capability of CMB maps to determine
these cosmological parameters.  The answer to this question will
depend on the measurement errors $\sigma_l$, and on
the underlying cosmological theory.  If the actual parameters
describing the Universe are ${\bf s}_0$, then
the probability distribution for observing a CMB power spectrum which
is best fit by the parameters ${\bf s}$ is
\begin{equation}
P({\bf s}) \propto \exp\left[ -{1\over 2}({\bf s}-{\bf s}_0)
                     \cdot [\alpha] \cdot({\bf s}-{\bf s}_0)\right]
\label{likelihood}
\end{equation}
where the curvature matrix $[\alpha]$ is given approximately by
\begin{equation}
\alpha_{ij} = \sum_\ell {1\over\sigma_\ell^2}
              \left[{\partial C_\ell({\bf s}_0)\over\partial s_i}
                    {\partial C_\ell({\bf s}_0)\over\partial s_j}\right]
\label{curvature}
\end{equation}
with $\sigma_\ell$ as given in Eq.~(\ref{variance}).
The covariance matrix $[{\cal C}] = [\alpha]^{-1}$ is an estimate of the
standard errors that would be obtained from a maximum-likelihood
fit to data: the variance in measuring the parameter $s_i$ (obtained by
integrating over all the other parameters) is
approximately ${\cal C}_{ii}^{1/2}$. If some of the parameters
are known, then the covariance matrix for the others is
determined by inverting the submatrix of the
undetermined parameters. For example, if
all parameters are fixed except for $s_i$, the variance in $s_i$
is simply $\alpha_{ii}^{-1/2}$.
In previous work, variances were estimated for small subsets of
the parameters with Monte Carlo calculations \cite{knox,hinshaw};
the present approach can be used to reproduce these results.

\begin{figure}[t]
\centerline{\psfig{figure=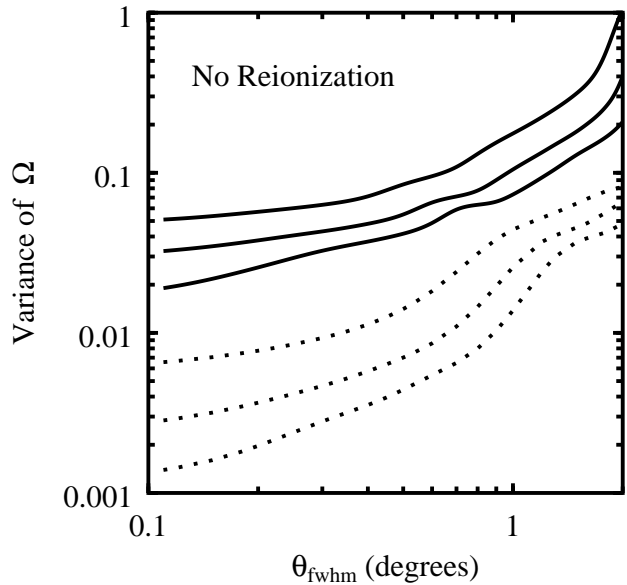,height=4in}}
\caption{The variance on $\Omega$ that can be obtained with a
     full-sky mapping experiment as a function of the beam width
     $\theta_{\rm fwhm}$ for noise levels $w^{-1}=2\times10^{-15}$,
     $9\times10^{-15}$, and $4\times10^{-14}$ (from lower to
     upper curves).  The underlying model is ``standard CDM.''
     The solid curves are the sensitivities attainable with no
     prior assumptions about the values of any of the other
     cosmological parameters.  The dotted curves are the
     sensitivities that would be attainable assuming that all
     other cosmological parameters, except the normalization,
     were fixed.  The results for
     a mapping experiment which covers only a fraction $f_{\rm
     sky}$ of the sky can be obtained by replacing $w
     \rightarrow w f_{\rm sky}$ and scaling by $f_{\rm
     sky}^{-1/2}$ [c.f., Eq.~(2)].}
\label{Omegafigure}
\end{figure}

Fig.~\ref{Omegafigure}\ displays the variance in $\Omega$ as a
function
of the beam width $\theta_{\rm fwhm}$ for different noise levels
and for $f_{\rm sky}=1$. For different values of $f_{\rm sky}$, replace
$w\rightarrow w f_{\rm sky}$ and scale by $f_{\rm sky}^{-1/2}$
[c.f., Eq.~(\ref{variance})].
The underlying model assumed here for the purpose of
illustration is ``standard CDM,'' given by ${\bf
s}=\{1,0.01,0.5,1,0,0,0,0,0,Q_{\rm COBE},3\}$, where $Q_{\rm
COBE}=20\,\mu$K is the COBE normalization \cite{cobenorm}.
The solid curves show the ${\cal C}_{\Omega\Omega}^{1/2}$ obtained by
inversion of the full $11\times11$ curvature matrix $[\alpha]$ for
$w^{-1}=2\times10^{-15}$, $9\times10^{-15}$, and $4\times
10^{-14}$.  These are the
sensitivities that can be attained at the given noise levels
with the assumption of uniform priors (that is, including {\it
no} information about any parameter values from other observations).
The dotted curves show the ${\cal C} _{\Omega\Omega}^{1/2}$
obtained by inversion of the $\Omega$-$Q$ submatrix of
$[\alpha]$; this is the
variance in $\Omega$ that could be obtained if all other
parameters except the normalization were fixed, either from
other observations or by assumption.  Realistically,
the precision obtained will fall somewhere between these
two sets of curves.  Other underlying models, including
low-$\Omega$ models, give similar sensitivities.
Although parameters other than $\Omega$
will have some weak effect on the position of the first Doppler
peak, they will also alter the
power spectrum at smaller angular scales.  Therefore, the
higher multipole moments accessible with smaller beam widths
help constrain the other parameters and
make the determination of $\Omega$ from the location of the
first Doppler peak more precise.

\begin{figure}[t]
\centerline{\psfig{figure=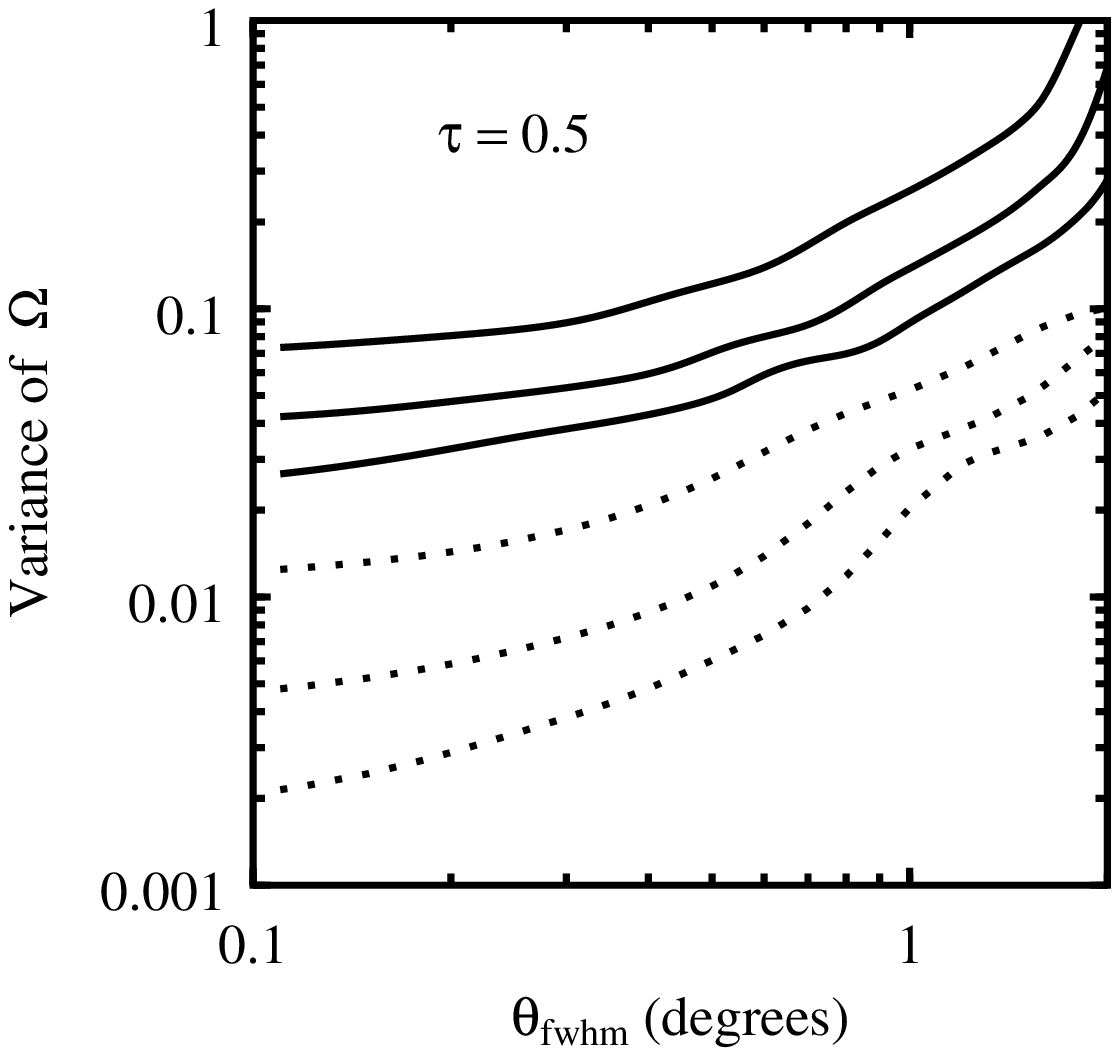,height=4in}}
\caption{Same as Fig.~1, except for a reionized model with
     $\tau=0.5$.}
\label{taufigure}
\end{figure}

Early reionization tends to wash out the structure of the
power-spectrum features, decreasing the precision of the
parameter estimates. To illustrate this effect, the curves in
Fig.~\ref{taufigure}\ show the same results as in
Fig.~\ref{Omegafigure}, but for a
reionized model in which $\tau = 0.5$.
As expected, the sensitivity to $\Omega$ decreases, although it
remains significant even for $\tau$ as large as one half.

At $\ell\gtrsim1000$, nonlinear effects become significant and
linear power-spectrum
calculations become unreliable.  Therefore, we extend the sum
in Eq.~(\ref{curvature}) only up to $\ell=1000$.
With improved calculations, the sensitivities at
small beam widths could conceivably be improved.

We have also investigated the sensitivity of CMB mapping
experiments to the other cosmological parameters listed
above.  Our results suggest that a map with $0.5^\circ$
angular resolution may also provide interesting constraints
to $\Lambda$ with minimal assumptions, and to the other parameters
with reasonable priors.   In particular, the experiments
should be able to distinguish between a flat matter-dominated
Universe and a flat cosmological-constant--dominated Universe.
These results will be presented
in detail elsewhere \cite{jkks}.

Figs.~\ref{Omegafigure}\ and \ref{taufigure}\ estimate the
probability of observing a set of
parameters given an underlying model. Actual data will require
solution of the inverse problem, estimating
the probability of an underlying
model given the data.  We are currently exploring the accuracy
with which all of the above parameters can be determined
given a simulated data set \cite{jkks,spergel}\ and to what extent
parameter degeneracy in current experiments can be resolved
\cite{confusion}.  Preliminary
results show that the true maximum of the likelihood function
can be recovered with good accuracy from a parameter search routine,
and that the errors in $\Omega$ approach the precision obtained here.

The numerical results presented here demonstrate that $\Omega$ can
be determined by realistic next-generation satellite experiments
with a precision on the order of a few percent. Such a
measurement will greatly solidify our knowledge of the gross
properties of the Universe, have crucial bearing on the
dark-matter and age problems, and will provide a
stringent test of the inflationary hypothesis.

\bigskip

We would like to thank Scott Dodelson, Lloyd Knox, Michael Turner,
and members of the MAP collaboration for helpful discussions.
This work is supported in part by the D.O.E. under contracts
DEFG02-92-ER 40699 at Columbia University
and DEFG02-85-ER 40231 at Syracuse University, by the Harvard
Society of Fellows, by the NSF under contract ASC 93-18185
(GC3 collaboration) at Princeton University,
and by NASA under contract NAGW-2448 and
under the MAP  Mission Concept Study Proposal at Princeton
University. M.K. acknowledges the hospitality of the NASA/Fermilab
Astrophysics Center where part of this work was completed.

\vfil\eject

\end{document}